\documentclass{TCAM}

\usepackage[english]{babel}    

\usepackage[utf8]{inputenc}    

\usepackage{epsfig}
\usepackage{multirow}
\usepackage{array}
\usepackage{url}
\usepackage{graphics}
\usepackage{subfigure}
\usepackage{psfrag}

\usepackage{framed}
\usepackage{tikz}

\usepackage{listings}
\usepackage{mathrsfs}
\usepackage{amsmath}
\usepackage{amstext}
\usepackage{amsfonts}

\DeclareMathOperator{\sech}{sech}

\title{The wave stability of solitary waves over a bump for the full Euler equations}

\author{ Marcelo V. Flamarion 
	\aff{UFRPE/Rural Federal University of Pernambuco, UACSA/Unidade Acad{\^e}mica do Cabo de Santo Agostinho, BR 101 Sul, 5225, 54503-900, Ponte dos Carvalhos, Cabo de Santo Agostinho, Pernambuco, Brazil -- E-mail:  marcelo.flamarion@ufrpe.br}, 
	\&  %
	Roberto Ribeiro-Jr%
	\aff{UFPR/Federal University of Paran{\' a}, Departamento de Matem{\' a}tica, Centro Polit{\' e}cnico, Jardim das Am{\' e}ricas, Caixa Postal 19081, Curitiba, PR, 81531-980, Brazil -- E-mail: robertoribeiro@ufpr.br }
	}\abstracttcam{In this work, we present a numerical study of the wave stability of steady solitary  waves over a localised topographic obstacle through the full Euler equations. There are two branches of the solutions: one from the perturbed uniform flow and the other from the perturbed solitary-wave flow. We find that steady waves from the perturbed uniform flow are always stable with respect to perturbations of its amplitude. Regarding the perturbed solitary-wave, when the perturbed initial condition has smaller amplitude than the steady solution we notice  a certain type of stability. Yet, when the perturbed initial condition has larger amplitude than the steady solution an onset of wave-breaking seem to occur. }
\keywords{Water waves, Conformal mapping, Euler equations, Wave stability}

\begin{document}
	
	\maketitle
	
\section{Introduction}

Water waves is a field of many interesting physical problems. For instance, problems related to the propagation of water waves over topographic obstacles \cite{Baines, Pratt}, ship wakes  and ocean waves generated by storms \cite{Johnson}.

The interaction wave-current-topography has been extensively studied in the past few years using  different mathematical models. Perhaps, the forced Korteweg-de Vries equation (fKdV) is the more  commonly nonlinear model used. The fKdV equation arises as a model for submerged obstacles with small amplitudes in  nearly-critical flows, i.e., when the  Froude number defined as
\begin{equation}
F = \frac{U_0}{\sqrt{gh_0}},
\end{equation}
is close to 1. Here, $U_0$ is the uniform flow speed, $g$ is the gravity and $h_0$ is the average depth of the channel. The flow is called supercritical or subcriticall depending on whether $F >1$ or $ F< 1$. A careful study on this model was first done by Wu and Wu \cite{Wu2} and later by several other authors \cite{Akylas, Grimshaw86, Wu1, Wu2, Paul, Marcelo-Paul-Andre, Capillary, M-Selecciones}. 

On the light of the full Euler equations,  Vanden-Broeck and Tuck \cite{Tuck} investigated steady subcritical waves generated by a moving pressure distribution and their connection with ship generated waves. Later, Asavanant et al. \cite{Asavanant} studied the same problem considering both the subcritical and supercritical regimes. They explored different parameter regimes, including effects of pressure intensity and distribution length. Binder et al. \cite{Binder} used the boundary integral method to compute steady supercritical solutions in the presence of two triangles along the bottom.

More recently, Grimshaw and Maleewong \cite{Grimshaw13} studied the stability of steady solutions of fKdV equation in both subcritical and supercritical regimes. They found their steady wave from the transient fKdV solution, whose stability was then analyzed through the Euler equations in the presence of a moving pressure distribution. In the presence of a constant current and a topography, Vanden-Broeck \cite{Broeck} used a boundary integral method to compute steady solutions for the Euler equations. Several types of steady waves were found but their stability was not analyzed. Later, Flamarion et al. \cite{Marcelo-Paul-Andre} presented an iterative numerical method based on conformal mapping technique to study waves generated by a current-topography interaction for the full Euler equations and compared their results with the ones produced by the fKdV equation. They observed that the two models agreed well in the weakly nonlinear weakly dispersive regime. In addition, a few types of steady waves were computed through a Newton's method type.

In this work, we compute numerically steady waves for the full Euler equations in the presence of a topographic obstacle and study their wave stability. Although other authors have already study the same problem, the novelty of the present work is the study of the wave stability. We find two branches of solutions: the perturbed uniform flow and the perturbed solitary-wave flow. The steady waves from the perturbed uniform flow are always stable with respect to perturbations of its amplitude. Regarding the perturbed solitary-wave, when the perturbed initial condition has smaller amplitude than the steady solution we find a certain type of stability. However, the steady wave solutions are unstable when the perturbed initial condition has larger amplitude than the steady solution and an onset of a wave-breaking seem to occur at later times. This study is a natural step up from fKdV results reported by Chardad et al. \cite{Chardad}.

This article is organized as follows. In section 2 we present the mathematical formulation of the Euler equations. In section 3 we describe the conformal mapping technique and rewrite the Euler equations in the canonical domain, which is a uniform strip. In section 4 we present a numerical method to solve them. Section 5 contains the numerical results and section 6 the conclusion.

\section{Mathematical Formulation}
We consider a two-dimensional incompressible and irrotational flow of an inviscid fluid with constant density $(\rho)$ in the presence of gravity $(g)$, a uniform upstream current ($U_0$) in the presence of a topographic obstacle $h(x)$ in a channel with typical depth $h_0$ in the far field. We denote the velocity potential by $\phi(x,y,t)$ and the free surface profile by $\bar{\zeta}(x,t)$. We choose $h_{0}$, $(gh_0)^{1/2}$ and $(h_{0}/g)^{1/2}$ as our reference units in space, speed and time, respectively.   
Thus, the dimensionless Euler equations are
\begin{align} \label{Eu1}
\begin{split}
& \overline{\phi}_{xx}+\overline{\phi}_{yy}= 0, \;\  \mbox{for} \;\ -1+ h(x) < y <\overline{\zeta}(x,t), \\
& Fh_{x} + \overline{\phi}_{x}h_{x} =\overline{\phi}_{y}, \;\ \mbox{at} \;\ y = -1+ h(x), \\
& \overline{\zeta}_{t}+F\overline{\zeta}_{x}+\overline{\phi}_{x}\overline{\zeta}_{x}-\overline{\phi}_{y}=0,
\;\ \mbox{at} \;\ y = \overline{\zeta}(x,t), \\
& \overline{\phi}_{t}+F\overline{\phi}_{x}+\frac{1}{2}(\overline{\phi}_{x}^2+\overline{\phi}_{y}^{2}) +\overline{\zeta}= 0, \;\ \mbox{at} \;\ y = \overline{\zeta}(x,t),
\end{split}
\end{align}
where $F ={U_0}/{(gh_0)^{1/2}}$ is the Froude number. 

In the next section we rewrite equations (\ref{Eu1}) using the conformal mapping technique, which allow us to solve them  numerically.

\section{Conformal mapping}
Consider the conformal mapping from the canonical $w$-plane ($w=\xi+ i \eta$) onto the physical $z$-plane ($z=x+i y$),
\begin{equation*}
z(\xi,\eta,t) = x(\xi,\eta,t)+iy(\xi,\eta,t),
\end{equation*}
satisfying the boundary conditions
\begin{equation*}
y(\xi,0,t)=\overline{\zeta}(x(\xi,0,t),t) \;\ \mbox{and} \;\ y(\xi,-D,t)=-1+\mathbf{H}(\xi,t),
\end{equation*}
where $\mathbf{H}(\xi,t) = h(x(\xi,-D,t))$. It is required that the canonical strip's height $D$ is a function of time $t$. 
 $D=D(t)$  depends on the wave profile and will be determined later.  We denote by $\phi(\xi,\eta,t) =\bar{\phi}(x(\xi,\eta,t),y(\xi,\eta,t),t)$ and $\psi(\xi,\eta,t)=\bar{\psi}(x(\xi,\eta,t),y(\xi,\eta,t),t)$ the potential and its harmonic conjugate in the canonical domain.   Let $\mathbf{\Phi}(\xi,t)$, $\mathbf{\Psi}(\xi,t)$,  $\mathbf{X}(\xi,t)$ and $\mathbf{Y}(\xi,t)$ be the traces of $\phi$, $\psi$, $x$ and $y$ at $\eta=0$, respectively. Substituting these variables in Kinematic and Bernoulli conditions $(\ref{Eu1})_{3,4}$ a straight-forward computation shows that the Euler equations in the canonical domain are
\begin{align}\label{Eu2}
\begin{split}
& \mathbf{X}_{\xi} = 1-\mathcal{C}\bigg[\mathbf{Y}_{\xi}-\mathcal{F}^{-1}\bigg( \frac{\widehat{\mathbf{H}}_{\xi}(k_j,t)}{\cosh(k_jD)}\bigg)\bigg], \\
& \mathbf{\Phi}_{\xi} = -\mathcal{C}\bigg[\mathbf{\Psi}_{\xi}(\xi,t)+\mathcal{F}^{-1}\bigg(\frac{F\widehat{\mathbf{H}}_{\xi}(k_j,t)}{\cosh(k_jD)}\bigg)\bigg], \\
& \mathbf{Y}_{t} =\mathbf{Y}_{\xi}\mathcal{C}\bigg[\frac{\Theta_{\xi}}{J}\bigg] 
-\mathbf{X}_{\xi}\frac{\Theta_{\xi}}{J}, \\
& \mathbf{\Phi}_{t} = - \mathbf{Y} - \frac{1}{2J}
(\mathbf{\Phi}_{\xi}^{2}-\mathbf{\Psi}_{\xi}^{2}) +\mathbf{\Phi}_{\xi}\mathcal{C}\bigg[\frac{\Theta_{\xi}}{J}\bigg] 
- \frac{1}{J}F\mathbf{X}_{\xi}\mathbf{\Phi}_{\xi}, \\
& \mathbf{X}_{b}(\xi,t) =x(\xi,-D,t) = \xi -\mathcal{C}\Bigg[\mathcal{F}^{-1}\Bigg(\frac{\widehat{\mathbf{Y}}(k_j,t)}{\cosh(k_jD)}-\frac{\widehat{\mathbf{H}}(k_j,t)}{\cosh^{2}(k_jD)}\Bigg)\Bigg ]+\mathcal{T}\Big[\mathbf{H}(k_j,t)\Big],
\end{split}
\end{align}
where   $\mathbf{\Theta}_{\xi}(\xi,t)=\mathbf{\Psi}_{\xi}+F\mathbf{Y}_{\xi}$, $J=\mathbf{X}_{\xi}^2+\mathbf{Y}_{\xi}^2$ is the Jacobian of the conformal mapping evaluated at $\eta=0$, $\mathcal{C}$ and $\mathcal{T}$  are the operators  $$\mathcal{C}=\mathcal{F}^{-1}_{k_j\ne 0}i\coth(k_jD)\mathcal{F}_{k_j\ne 0} \mbox{ and } \mathcal{T}=\mathcal{F}^{-1}_{k_j\ne 0}i\tanh(k_jD)\mathcal{F}_{k_j\ne 0},$$ 
where the Fourier modes are given by $$\mathcal{F}_{k_j}[g(\xi)]=\hat{g}(k_j)=\frac{1}{2L}\int_{-L}^{L}g(\xi)e^{-ik_j\xi}\,d\xi,$$
$$\mathcal{F}^{-1}_{k_j}[\hat{g}(k_j)](\xi)=g(\xi)=\sum_{j=-\infty}^{\infty}\hat{g}(k_j)e^{ik_j\xi},$$
with $k_j=(\pi/L)j$, $j\in\mathbb{Z}$. According to our formulation $2L$ is the length of the  canonical domain. By imposing the physical and canonical domain to have the same length we find that  $$D(t) = 1+ \frac{1}{2L}\int_{-L}^{L}\mathbf{Y}(\xi,t)-\mathbf{H}(\xi,t)d\xi.$$
More details of this conformal mapping are presented in \cite{Dyachenko, Marcelo-Paul-Andre, International}.

Steady waves are obtained from the set of equations (\ref{Eu2}) imposing  $\partial_{t}=0$. Following \cite{Marcelo-Paul-Andre} we conclude that 
$\mathbf{\Theta}=0$ and 
\begin{equation*}
\mathbf{\Psi}_{\xi}(\xi) = -F\mathbf{Y}_{\xi}(\xi).
\end{equation*}
Therefore, equations (\ref{Eu2}) are now written as
\begin{align}\label{Eus}
\begin{split}
& \mathbf{X}_{\xi}(\xi) = 1-\mathcal{C}\bigg[\mathbf{Y}_{\xi}-\mathcal{F}^{-1}\bigg( \frac{\widehat{\mathbf{H}}_{\xi}(k_j)}{\cosh(k_jD)}\bigg)\bigg], \\
& \mathbf{\Phi}_{\xi}(\xi) = -\mathcal{C}\bigg[\bigg(\mathbf{\Psi}_{\xi}(k_j)+\mathcal{F}^{-1}\bigg(\frac{F\widehat{\mathbf{H}}_{\xi}(k_j)}{\cosh(k_jD)}\bigg)\bigg)\bigg], \\
& \mathbf{Y} + \frac{1}{2J}
(\mathbf{\Phi}_{\xi}^{2}-\mathbf{\Psi}_{\xi}^{2}) + \frac{1}{J}F\mathbf{X}_{\xi}\mathbf{\Phi}_{\xi} =0, \\
& \mathbf{X}_{b}(\xi) =x(\xi,-D) = \xi -\mathcal{C}\Bigg[\mathcal{F}^{-1}\Bigg(\frac{\widehat{\mathbf{Y}}(k_j)}{\cosh(k_jD)}-\frac{\widehat{\mathbf{H}}(k_j)}{\cosh^{2}(k_jD)}\Bigg)\Bigg ]+\mathcal{T}\Big[\mathbf{H}(k_j)\Big],
\end{split}
\end{align}

In the next section, we present the numerical methods to compute steady waves and their evolution.

\section{Numerical Methods}
The numerical approachs presented bellow is the same reported in \cite{Marcelo-Paul-Andre}. Here, we only summarise the main steps.

\subsection{Steady wave solutions}
Numerical steady waves of the Euler equations (\ref{Eus}) are found on a domain $\xi\in [-L,L]$, with $N$ uniformily spaced points with grid size $\Delta\xi=2L/N$. On the grid points $\xi_n$, $n = 1,2,...N,$ the free surface elevation  is denoted by $Y_n=\mathbf{Y}(\xi_n)$. The steady Bernoulli equation is written as
\begin{align}\label{EE3}
\begin{split}
G_{n}({Y}_1,{Y}_2,...,{Y}_{N}):= \mathbf{Y}(\xi_n) + \frac{1}{2J}
(\mathbf{\Phi}_{\xi, n}^{2}-\mathbf{\Psi}_{\xi, n}^{2}) + \frac{1}{J}F\mathbf{X}_{\xi, n}\mathbf{\Phi}_{\xi, n}=0.
\end{split}
\end{align}
Fourier transforms and the operator $\mathcal{C}$  are approximated by the FFT on the uniform grid, and all derivatives are performed spectrally \cite{Trefethen}. The Jacobian for 
Newton's method is computed using 
\begin{align}\label{G}
\begin{split}
\frac{\partial G_n}{\partial{Y}_l}=\frac{G_{n}({Y}_1,{Y}_2,...,{Y}_l+\delta,...,{Y}_{N})-G_{n}({Y}_1,{Y}_2,...,{Y}_l,...,{Y}_{N})}{\delta},
\end{split}
\end{align}
and the stopping criteria for the Newton's method is
\begin{align*}
\begin{split}
\frac{\sum_{j=1}^{N}|G_{n}({Y}_1,{Y}_2,...,{Y}_{N})|}{N}< 10^{-16}.
\end{split}
\end{align*}
The topography $\mathbf{H}(\xi)$ is computed iteratively   by solving 
\begin{align}\label{scheme}
	\begin{split}
		& \mathbf{X}_{b}^{l}(\xi)= \xi -\mathcal{C}\Bigg[\mathcal{F}^{-1}\Bigg(\frac{\widehat{\mathbf{Y}}(k_j)}{\cosh(k_jD)}-\frac{\widehat{\mathbf{H}}^l(k_j)}{\cosh^{2}(k_jD)}\Bigg)\Bigg ]+\mathcal{T}\Big[\mathbf{H}^l(k_j)\Big],  \\
		& \mathbf{H}^{l+1}(\xi)=h(\mathbf{X}_{b}^{l}(\xi)).
	\end{split}
\end{align}
The initial step is  $\mathbf{X}_{b}^{0}(\xi)=\xi$ and $\mathbf{H}^{0}(\xi)=h(\xi)$. The scheme is performed with  the stopping criteria
$$\frac{\displaystyle\max_{\xi\in[-L,L]}\Big|\mathbf{H}^{l+1}(\xi)-\mathbf{H}^{l}(\xi)\Big|}{\displaystyle\max_{\xi\in[-L,L]}\Big|\mathbf{H}^{l}(\xi)\Big|}<10^{-16}.$$

\subsubsection{Initial guess for the Newton's method and topography's profile}
We are interested in studying steady solitary waves solutions for the full Euler equations.

It is well known that, in the nearly-critical regime ($F=1+\epsilon f$, where $f$ is a small parameter)  and for obstacles of small amplitudes, the forced Korteweg-de Vries equation 
\begin{equation}\label{fKdV}
\bar{\zeta}_{t}+f\bar{\zeta}_{x}-\frac{3}{2}\bar{\zeta}\bar{\zeta}_{x}-\frac{1}{6}\bar{\zeta}_{xxx}=\frac{1}{2}h_{x}(x),
\end{equation}
can be obtained asymptotically from equations (\ref{Eu1}) its solutions agree well with the solutions of the Euler equations \cite{Marcelo-Paul-Andre}. This motivates us to use steady wave solutions of fKdV as initial guess of the Newton's method.  To this end, we proceed in the same fashion as presented in \cite{Chardad}. We impose
\begin{equation}
\bar{\zeta}(x) = A\sech^{2}(\beta x),
\end{equation}
to be a steady solution of (\ref{fKdV}). Thus, the topography satisfies 
$$h(x) = 2f\bar{\zeta} -\frac{3}{2}\bar{\zeta}^{2}-\frac{1}{3}\bar{\zeta}_{xx}.$$
In other words,
\begin{equation}\label{Top}
h(x) = \frac{A}{6}\Bigg(\frac{12f-8\beta^{2}}{\cosh^{2}(\beta x)}+\frac{12\beta^{2}-9A}{\cosh^{4}(\beta x)}\Bigg).
\end{equation}
Choosing the topographic obstacle to be a $\sech^{2}$-type we obtain the two branch of solutions
\begin{align}\label{solution}
	\begin{split}
		& A = f\pm \sqrt{f^{2}-G}, \\
		& \beta=\sqrt{\frac{3A}{4}},
	\end{split}
\end{align}
where $G$ is the amplitude of the obstacle. Camassa and Wu \cite{Camassa1, Camassa2} showed that the perturbed solitary-wave solution with $A = f+ \sqrt{f^{2}-G}$ is always unstable. On the other hand the perturbed uniform flow solution $A = f- \sqrt{f^{2}-G}$ is stable only if $G\le\frac{80}{81}f^{2}$.

The fKdV variables are related to the Euler's ones according to the transformations
\begin{equation}\label{scale1}
x\rightarrow \epsilon^{1/2}x, \;\ t\rightarrow  \epsilon^{3/2}t, \;\ \bar{\zeta}\rightarrow\epsilon \bar{\zeta}, h\rightarrow\epsilon^{-2}h \mbox{ and }  F=1+\epsilon f.
\end{equation}

We set the topographic obstacle for the Euler equations to be the rescaled topography of the fKdV (\ref{Top})
\begin{equation}\label{TopographySteady}
h(x) = \epsilon^{2}G\sech^{2}(\epsilon^{1/2}\beta x),
\end{equation}
where $\epsilon>0$ is a small parameter.

For this choice of topography, it is natural to consider as initial guess for the Newton's method 
\begin{equation}\label{Guess}
\mathbf{Y}(\xi) = \epsilon A\sech^{2}(\epsilon^{1/2}\beta\xi),
\end{equation}
where $A$ and $\beta$ are determined by equation (\ref{solution}) with $f=0.32$ and $G=0.1$. These solutions are then continued in the parameter $f$ using the Newton's Method continuation, which allow us to determine two branch of solutions:  solutions of the perturbed solitary-wave and solutions of the perturbed uniform flow. 

\subsection{Time-dependent wave solutions}
The evolution of the initial data of equations (\ref{Eu2}) is found by integrating in time the family of ordinary differential equations
through the fourth-order Runge--Kutta method and the derivatives in $\xi$ are performed using the Fast Fourier Transform (FFT) \cite{Trefethen}. The topography $\mathbf{H}(\xi,t_m)$ at time $t=t_m$ is computed iteratively   by solving 
\begin{align}\label{scheme}
	\begin{split}
		& \mathbf{X}_{b}^{l+1}(\xi,t_m)= \xi -\mathcal{C}\Bigg[\mathcal{F}^{-1}\Bigg(\frac{\widehat{\mathbf{Y}}(k_j,t_m)}{\cosh(k_jD)}-\frac{\widehat{\mathbf{H}}^l(k_j,t_m)}{\cosh^{2}(k_jD)}\Bigg)\Bigg ]+\mathcal{T}\Big[\mathbf{H}^l(k_j,t_m)\Big],  \\
		& \mathbf{H}^{l+1}(\xi,t_m)=h(\mathbf{X}_{b}^{l}(\xi,t_m)),
	\end{split}
\end{align}
for $l\ge 0$. Equation (\ref{scheme}) is solved using as  initial step  $\mathbf{X}_{b}^{0}(\xi,t_m)=\xi$ and $\mathbf{H}^{0}(\xi,t_m)=h(\xi)$. The scheme is performed with  the stopping criteria
$$\frac{\displaystyle\max_{\xi\in[-L,L]}\Big|\mathbf{H}^{l+1}(\xi,t_m)-\mathbf{H}^{l}(\xi,t_m)\Big|}{\displaystyle\max_{\xi\in[-L,L]}\Big|\mathbf{H}^{l}(\xi,t_m)\Big|}<10^{-16}.$$

\section{Numerical Results}

In this section we compute steady solutions using different values of $\epsilon$. The branches of steady solutions is compared with the ones of the fKdV equation.  We then perturb the initial data (steady wave) and compute its evolution numerically in order to investigate wether these solutions are stable. 

For the fKdV equation, steady waves and their related stability properties were studied in  \cite{Camassa1, Camassa2, Chardad}.
More recently, Grimshaw and Maleewong \cite{Grimshaw13} analysed the stability of steady fKdV solutions in both the subcritical ($F<1$) and supercritical ($F>1$) regimes. 
They found steady waves from the transient fKdV solution, whose stability was then analysed through    
the Euler equations in the presence of a moving pressure distribution. In the presence of a constant current and a topography, Vanden-Broeck 
 \cite{Broeck} used a boundary integral method to compute steady solutions to the Euler equations. Different steady waves were found, but their stability was not analysed numerically.
\begin{figure}[h!] 
\centering
\includegraphics[scale=1]{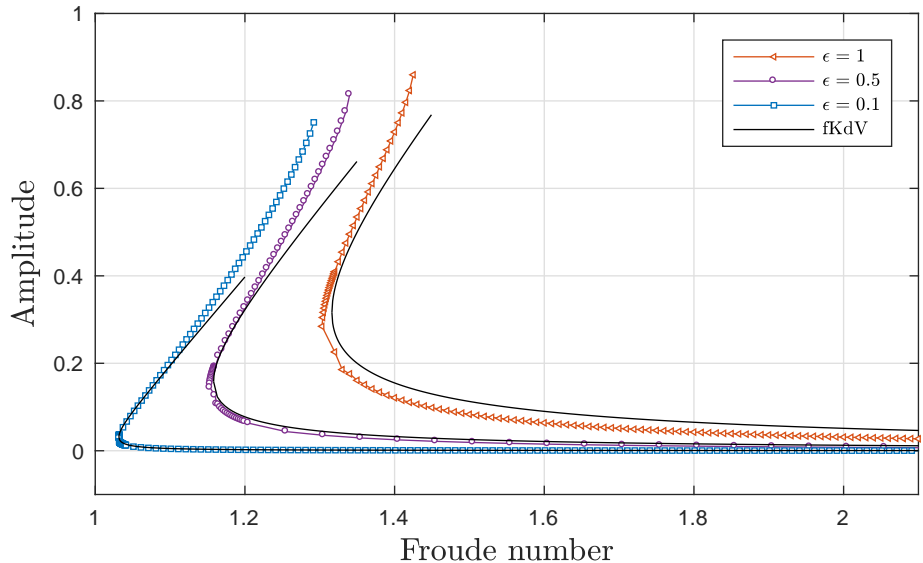} 
\caption{Comparison between the the branch of steady solutions of the full Euler equations for different values of the parameter $\epsilon$ and the branch of steady solutions of the fKdV model.}
\label{Fig1}
\end{figure}  
 
Initially, we compute steady solutions using different values of $\epsilon$ through the numerical method described in the previous section. When $\epsilon$ approaches zero and the Froude number is nearly-critical ($F\approx 1$), the two branches (the uniform flow and the solitary-wave) of steady solutions of the full nonlinear model is close to the one predicted by the weakly nonlinear weakly dispersive theory. However, as the Froude number increases the solitary-wave solutions no longer agree with the solutions of the fKdV model, which does not occur in the uniform flow solutions. More details are given in Figure \ref{Fig1}. As it can be seen, as we allow for a gradual increase of the topography's amplitude, the solutions of the two models start to differ and the nonlinear theory predicts solutions of the  solitary-wave branch with higher amplitude.

Now, we investigate the wave stability of the steady waves computed through the Newton's method type by disturbing its initial amplitude and setting it as an initial data for the time-dependent Euler equations (\ref{Eu1}). It is natural to expect that for small values of $\epsilon$, the results produced by performing numerical simulations with the full nonlinear model to be similar to the ones reported using the fKdV model -- for instance see Chardard et al. \cite{Chardad}.  
\begin{figure}[h!]
\centering
\includegraphics[scale=1]{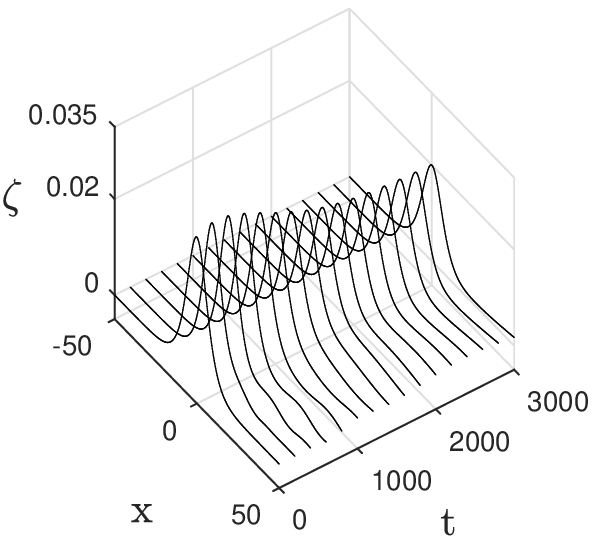} 
\includegraphics[scale=1]{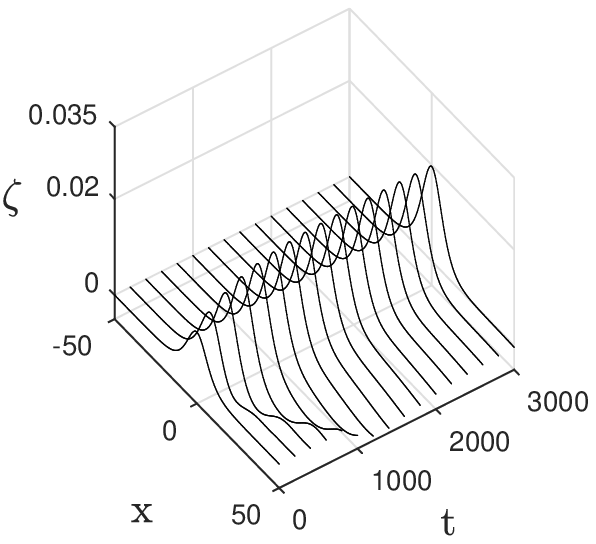} 
\caption{The evolution of the stable perturbed uniform flow solution of the Euler equations when $\epsilon=0.1$ and $F = 1.035$. The initial condition is the steady wave times $1.5$ (left) and $0.5$ (right).}
\label{Fig2}
\end{figure}

Figure \ref{Fig2} displays the evolution of a perturbed solution of the uniform flow with $\epsilon=0.1$. The solution is stable in the sense that, when its amplitude is perturbed, the numerical solution tend to recover its natural steady state (the decrease in amplitude of the wave on the right of Figure \ref{Fig2} and the increase in amplitude of the wave on the left of Figure \ref{Fig2}). 
\begin{figure}[h!]
\centering
\includegraphics[scale=1]{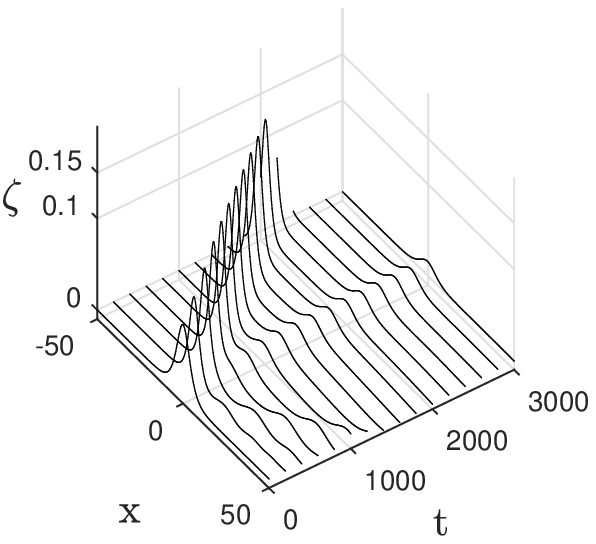} 
\includegraphics[scale=1]{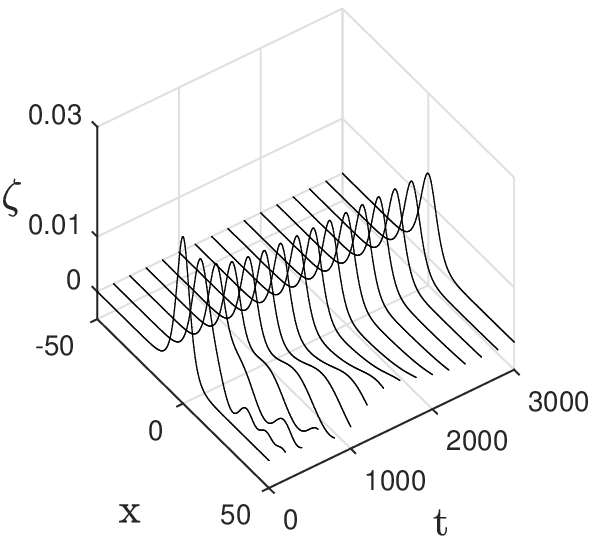} 
\caption{The evolution of the unstable perturbed solitary-wave solution of the Euler equations when $\epsilon=0.1$ and $F = 1.035$. The initial condition is the steady wave times $1.5$ (left) and $0.5$ (right).}
\label{Fig3}
\end{figure}

Differently from the perturbed uniform flow, the perturbed solitary-wave does not recover its initial state after disturbing its initial amplitude. However, the system is somehow stable in the sense that a perturbed uniform flow solution arises above the obstacle for large times. When the perturbed initial condition has amplitude  smaller than the steady solution, its amplitude decreases in time and approaches to the perturbed uniform flow solution. On the other hand, when the perturbed initial condition has larger amplitude than the steady solution, a large solitary wave propagates upstream, leaving behind a small steady wave, which happens to be the perturbed uniform flow solution. These behaviours are depicted in Figure \ref{Fig3}.
\begin{figure}[h!]
\centering
\includegraphics[scale=1]{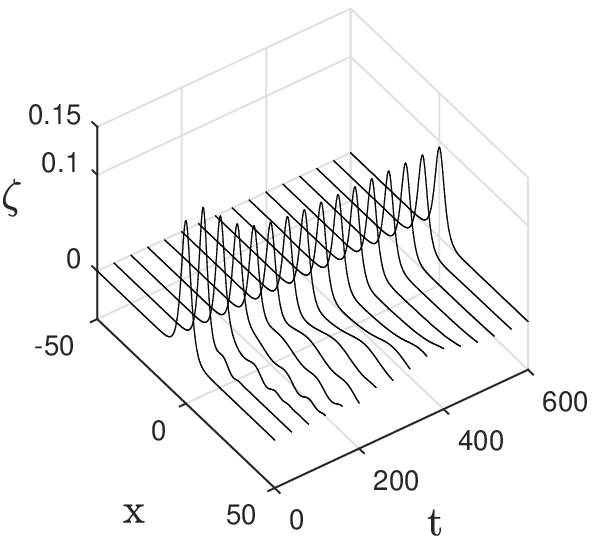} 
\includegraphics[scale=1]{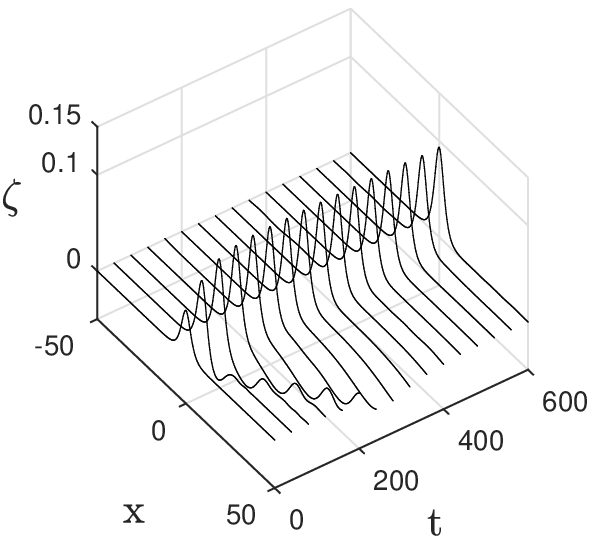} 
\caption{The evolution of the stable perturbed uniform flow solution of the Euler equations when $\epsilon=0.5$ and $F = 1.175$. The initial condition is the steady wave times $1.5$ (left) and $0.5$ (right).}
\label{Fig4}
\end{figure}

Next, we allow the amplitude of the topographic obstacle increases. Fixing $\epsilon=0.5$, the perturbed solution of the uniform flow turns out to be stable for small perturbation in the amplitude. A typical example is depicted in Figure \ref{Fig4}. The behaviour is qualitatively similar to the one predicted by the fKdV model. 
\begin{figure}[h!]
\centering
\includegraphics[scale=1]{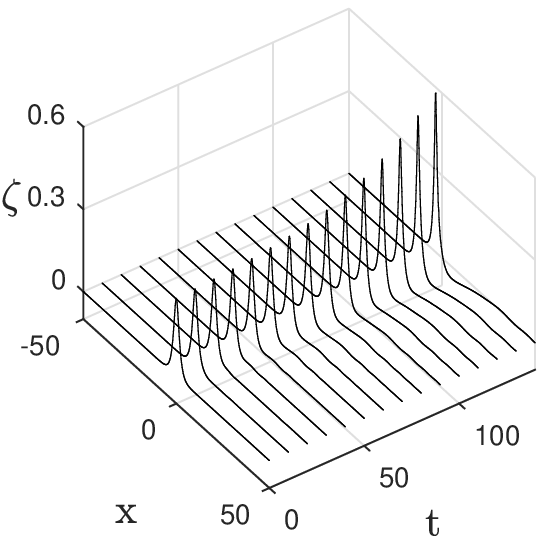} 
\includegraphics[scale=1]{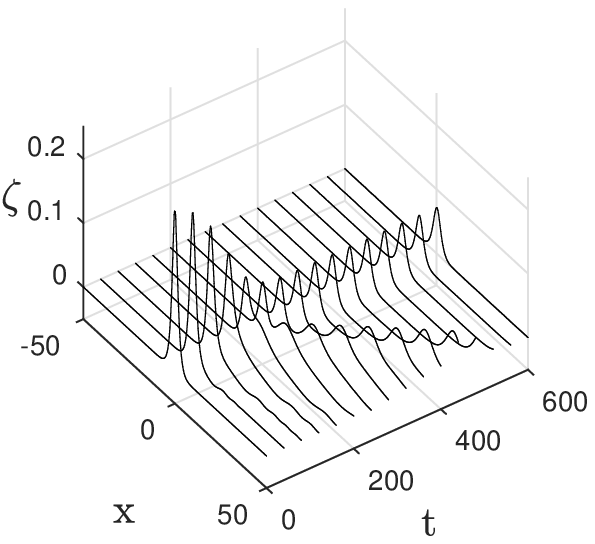} 
\caption{The evolution of the unstable perturbed solitary-wave solution of the Euler equations when $\epsilon=0.5$ and $F = 1.175$. The initial condition is the exact solution times $1.05$ (left) and $0.95$ (right).}
\label{Fig5}
\end{figure}

Regarding the solitary-wave perturbed solutions, we see that the weakly nonlinear weakly dispersive model is no longer appropriate to study steady waves, for instance $\epsilon=0.5$.  When the perturbed initial condition has amplitude  smaller than the steady solution, its amplitude decreases in time and approaches to the perturbed uniform flow solution. However, when the perturbed initial condition has amplitude  larger than the steady solution, its amplitude increases with time towards to a value which indicates that this wave may break. This behaviour is similar to the ones reported in the works of Grimshaw and Maleewong \cite{Grimshaw13} and Flamarion et al. \cite{Marcelo-Paul-Andre} in the context of generated waves by moving disturbances.

\section{Conclusions}
In this article, we have presented a numerical method to study the wave stability of steady solitary water waves over an uneven topography using a conformal mapping.  We showed that the solitary waves perturbed from the uniform flow are always stable while  the ones from the perturbed solitary-wave present a certain type of stability when the perturbed initial condition has smaller amplitude and are unstable when the perturbed initial condition has larger amplitude than the steady solution.  Besides, we noticed that in the previous case  an onset of wave-breaking might occur.

%
%

\end{document}